\documentclass[twocolumn,aip,reprint,numerical]{revtex4-2}
\usepackage{array}
\usepackage{float}
\usepackage{textcomp}
\usepackage{bm}
\usepackage{tipa}
\usepackage{amsmath}
\usepackage{amssymb}
\usepackage{graphicx}
\usepackage{color}
\usepackage{lineno}
\usepackage{hyperref}
\usepackage{natbib}
\hypersetup{hypertex=true,
            colorlinks=true,
            linkcolor=blue,
            anchorcolor=blue,
            citecolor=blue}
\makeatletter

\newcommand{\lyxmathsym}[1]{\ifmmode\begingroup\def\b@ld{bold}
        \text{\ifx\math@version\b@ld\bfseries\fi#1}\endgroup\else#1\fi}

\usepackage{dcolumn}
\usepackage{bm}
\usepackage{epsfig}
\usepackage{braket}

\usepackage{lineno}
\makeatother

\begin{document}

\title{Optical properties of two-dimensional Dirac-Weyl materials with a flat band}

\date{\today}

\author{Li-Li Ye}
\affiliation{School of Electrical, Computer and Energy Engineering, Arizona State University, Tempe, AZ 85287, USA}

\author{Chen-Di Han}
\affiliation{School of Electrical, Computer and Energy Engineering, Arizona State University, Tempe, AZ 85287, USA}

\author{Ying-Cheng Lai} \email{Ying-Cheng.Lai@asu.edu}
\affiliation{School of Electrical, Computer and Energy Engineering, Arizona State University, Tempe, AZ 85287, USA}
\affiliation{Department of Physics, Arizona State University, Tempe, Arizona 85287, USA}

\begin{abstract}

The emergence of a flat band in Dirac-Weyl materials offers new possibilities for electronic transitions, leading to stronger interaction with light. As a result, the optical conductivity can be significantly enhanced in these flat-band materials as compared with graphene, making them potentially better candidates for optical sensing and modulation. Recently, a comprehensive theory for the optical conductivity of a spectrum of flat-band Dirac-Weyl materials has been developed, with explicit formulas for both the real and imaginary parts of the conductivity derived through two independent approaches. This Perspective offers a review of the development. An understanding of the optical properties of the flat-band Dirac-Weyl materials paves the way for optical device applications in the terahertz-frequency domain.

\end{abstract}

\maketitle

\section{Introduction} \label{sec:intro}

A frontier area of research in applied physics is two-dimensional (2D) Dirac-Weyl 
materials whose energy band consists of a pair of Dirac cones and a topologically 
flat band, electronic or optical~\cite{Sutherland:1986,BUGH:2009,SSWX:2010,Green2010,Dora2011,wang2011nearly,Huang2011,Mei:2012,Moitra2013,Raoux2014,GMBRWNSSV:2014,romhanyi2015hall,giovannetti2015kekule,Li2015,MSCGOAT:2015,VCMRMWSM:2015,TOINNT:2015,FZLC:2016,DLKDD:2016,Zhu2016,Bradlynaaf2016,Fulga2017,Ezawa2017,Zhong2017,Zhu2017,Drost2017,slot2017experimental,Tan2018}.
A flat band can also arise in metal-organic and covalent-organic 
materials~\cite{JNL:2021,NLLB:2022}.
This Perspective focuses on the optical properties but with a brief review of a 
number of phenomena related to the electronic and magnetic properties of these
materials.

To develop optical sensors and modulators based on Dirac-Weyl materials, the problem 
of adequate optical absorption must be addressed. Graphene, due to its linear 
dispersion relationship, has potential applications in developing optical 
devices~\cite{vakil2011transformation,grigorenko2012graphene,bao2012graphene}.
For example, graphene-based polarizers have the ability to select light 
polarization in a broad frequency range~\cite{bao2011broadband}. Graphene can 
also support high-frequency plasmon modes with frequency ranging from several 
terahertz to infrared, making it appealing for high-frequency 
communications~\cite{ju2011graphene,akyildiz2014terahertz} and ultrafast 
transform optics~\cite{li2014ultrafast,baudisch2018ultrafast}. The ability to 
generate strong polarization of light implies that materials coated with 
graphene can be exploited for applications at the two extremes: cloaking or
superscattering~\cite{chen2011atomically,li2015tunable,HL:2022b}. A difficulty in
such applications is that the light absorption rate of a single-layer graphene
is quite low - only a few percent. To significantly enhance the optical
absorption has attracted a great deal of interest since the beginning of the
field of 2D Dirac-Weyl materials. For example, it was found
that~\cite{thongrattanasiri2012complete,meng2020ultrathin}, when graphene is
coupled with a proper dielectric material, surface plasmon mode can arise so that
the achievable optical absorption rate can be over $90\%$. Such surface plasmon 
can also propagate in a graphene lattice with frequency above the terahertz
domain~\cite{mikhailov2007new,hanson2008dyadic,nikitin2011edge}, implying
potential applications in high frequency communication devices. It was also 
found that, in slightly twisted bilayer graphene, unusual plasmon modes and strong 
optical absorption can arise~\cite{ni2015plasmons,lin2020chiral,deng2020strong}.

In the study of the optical properties of graphene, the conventional way was to
treat the material as a thin layer with electric conductivity depending on the
angular frequency $\omega$ of the incident field, leading to the optical
conductivity $\sigma(\omega)$ that is typically complex~\cite{falkovsky2007space}. 
When the energy of the incident photon is below the Fermi energy $\mu$: 
$\hbar\omega < \mu$, only the intraband electron transition (from the conduction 
band to itself) is allowed. Such a process usually occurs for large devices in 
the frequency range of subterahertz and terahertz 
($0.1$-$10$ THz)~\cite{gao2014analytical,meng2020ultrathin}. For incident wave 
with a higher frequency, e.g., $\hbar\omega \approx 2\mu$, intraband transitions 
become insignificant and interband transitions from the valence to the conduction 
band dominate. For smaller devices, the optical field can be in the infrared to 
visible range~\cite{stauber2008optical,christensen2015localized}. 
Graphene plasmons are tunable by changing the Fermi energy, but the plasmon density
is frequency-dependent due to the different carrier densities at different
frequencies. A 1D topological electride with density-independent frequency was 
reported~\cite{WSGDLH:2019}. The simulation result was further verified by 
first-principle calculations on
$\text{Ba}_3\text{Cr}\text{N}_3$ and $\text{Sr}_3\text{Cr}\text{N}_3$.
Density-independent plasmons were predicted to arise in both 2D nodal line and 1D 
nodal point systems and confirmed by first-principle calculations~\cite{WSDLH:2019}.
In general, to fully characterize the electromagnetic properties of the material, 
both the real and imaginary parts of the optical conductivity are required.

In a recent work~\cite{HL:2022}, a comprehensive theory for the optical 
conductivity of a spectrum of 2D Dirac-Weyl materials was developed. It is the
so-called $\alpha$-$\mathcal{T}_3$ lattice system with graphene sitting at one 
end and pseudospin-1 material at the other end of the 
spectrum~\cite{BUGH:2009}. An $\alpha$-$\mathcal{T}_3$ lattice is formed
from the honeycomb graphene lattice by adding an extra atom at the center of 
each hexagonal unit cell~\cite{BUGH:2009}, with the normalized coupling strength
$\alpha t$ between this atom and any nearest neighboring atom in the cell, where 
$0\le \alpha \le 1$ and $t$ is the nearest-neighbor hopping energy in the original 
graphene lattice. The low energy excitations of the $\alpha$-$\mathcal{T}_3$ 
lattice can be described by the generalized Dirac-Weyl 
equation~\cite{BUGH:2009,XL:2016}, where the spinor wave function has three 
components. The lattice degenerates to graphene with pseudospin-1/2 quasiparticles 
for $\alpha=0$ - only in this limiting case is a flat band absent. For $\alpha > 0$, 
a flat band through the conic interaction of the two Dirac cones 
exists~\cite{Raoux2014,illes2015hall}. Under a continuum approximation, an 
$\alpha$-$\mathcal{T}_3$ lattice is effectively a thin conducting layer. Because 
of the flat band, three types of band-to-band transitions can occur: intraband, 
cone-to-cone, and flat-band-to-cone transitions. A general finding was that the 
extra transitions brought upon by the flat band can enhance the optical 
conductivity~\cite{HL:2022}.

Experimentally, photonic crystals can be used to generate $\alpha$-$\mathcal{T}_3$ 
lattices~\cite{RCF:2006,slot2017experimental,LF:2018,MDOTG:2018}. Electronically,
candidate materials include transition-metal oxide 
$\text{SrTiO}_3/\text{SrIrO}_3/\text{SrTiO}_3$ trilayer
heterostructures~\cite{wang2011nearly},
$\text{SrCu}_2(\text{BO}_3)_2$~\cite{romhanyi2015hall} or
graphene-$\text{In}_2\text{Te}_2$~\cite{giovannetti2015kekule}. 
Realization of other flat-band lattice systems is also 
possible~\cite{leykam2018artificial,franchina2020engineering}.

This Perspective is organized, as follows. In Sec.~\ref{sec:lattice}, several 
experimental lattice systems of 2D Dirac-Weyl flat-band materials are introduced.
The full optical-conductivity theory of these materials is reviewed in 
Sec.~\ref{sec:optical_conductivity}. Opinion on potential future research is offered in 
Sec.~\ref{sec:discussion}.

\section{Experimentally accessible lattice systems of 2D Dirac-Weyl flat-band materials} \label{sec:lattice}

\begin{figure} [ht!]
\centering
\includegraphics[width=\linewidth]{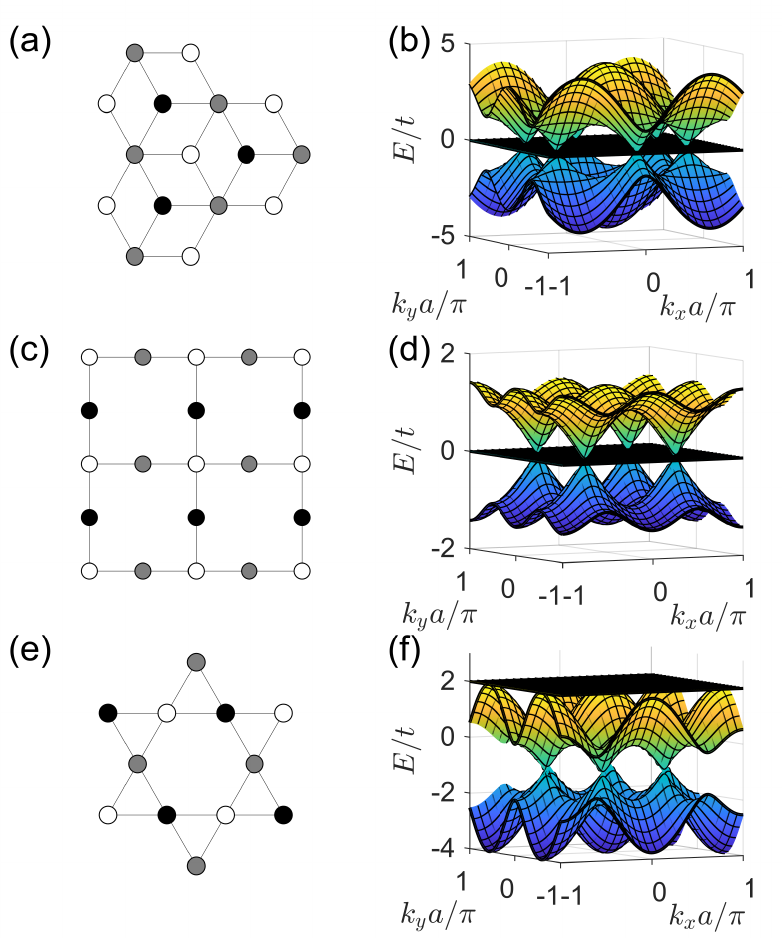}
\caption{Schematic illustration of three different lattice structures of 2D 
Dirac-Weyl materials with a flat band. (a,b) Dice lattice and its band structure, 
respectively. The first Brillouin zone has a honeycomb structure and two distinct 
valleys. (c,d) Lieb lattice and its band structure, respectively. The first 
Brillouin zone is a square with only one valley. (e,f) Kagome lattice and its
band structure, respectively. The first Brillouin zone is a honeycomb with the
same orientation as the original lattice, but the flat band arises at the top 
of the conduction band.} 
\label{fig:3_lattices}
\end{figure}

Figure~\ref{fig:3_lattices} shows three commonly studied lattice structures of 2D 
Dirac-Weyl flat-band materials - dice, Lieb, and Kagome lattices, together with their 
corresponding energy-band structures. The details of these lattices are described 
below. 

\subsection{Dice lattice}\label{subsec:Dice}

Dice lattice was originally proposed to study the Green's function for diatomic 
lattice systems~\cite{HC:1974}. The lattice is constructed by removing some 
couplings from a triangular lattice. The emergence of a flat band and a localization 
phenomenon in the dice lattice were first report in 
Ref.~[\onlinecite{Sutherland:1986}]. The localization behavior was later found to 
persist in the dice lattice system in the presence of a magnetic 
field~\cite{VMD:1998}. A similar phenomenon was also reported in quantum 
networks~\cite{VMD:2000}, in systems with spin-orbit coupling~\cite{BGCR:2005}, 
and in a Bose-Hubbard model~\cite{RCF:2006}. About 14 years ago, interest in the 
dice lattice was rejuvenated due to its unique structure of a pair of Dirac cones 
and a flat band~\cite{BUGH:2009}.

\begin{figure} [ht!]
\centering
\includegraphics[width=\linewidth]{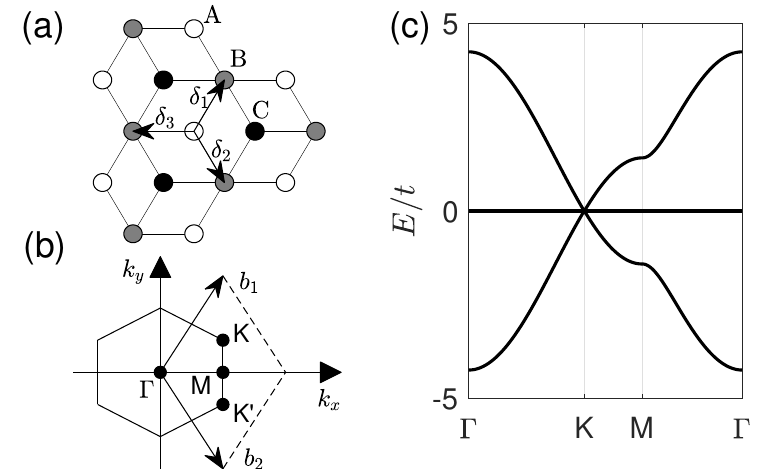}
\caption{A more detailed view of dice lattice and its band structure. 
(a) Dice lattice and the three nonequivalent atoms. The lattice unit vectors
are $\delta_i$ ($i=1,2,3$). (b) The first Brillouin zone. There are two 
nonequivalent Dirac points: $\mathbf{K}$ and $\mathbf{K}'$. (c) The band 
structure, where the conduction and valence bands as well as a flat band touch
each other at a Dirac point.}
\label{fig:S2_figure1}
\end{figure}

Dice lattice has three nonequivalent atoms, as shown in Fig.~\ref{fig:S2_figure1}(a).
The lattice has a three fold rotational symmetry and is also called the 
$\mathcal{T}_3$ lattice. The tight binding Hamiltonian describing the electronic 
structure of the dice lattice, taking into account nearest-neighbor hopping, is 
given by
\begin{align} \label{eq:S2_dice}
\begin{split}
\mathcal{H}(\mathbf{k})=&-t(e^{i\mathbf{k}\cdot \bm{\delta}}a_\mathbf{k}^\dagger b_\mathbf{k}+e^{-i\mathbf{k}\cdot\bm{\delta}}b_\mathbf{k}^\dagger a_\mathbf{k} \\
&+e^{i\mathbf{k}\cdot\bm{\delta}}b_\mathbf{k}^\dagger c_\mathbf{k}+e^{-i\mathbf{k}\cdot\bm{\delta}}c_\mathbf{k}^\dagger b_\mathbf{k})
\end{split}
\end{align}
where $t$ is the nearest-neighbor hopping energy, $a^\dagger,b^\dagger,c^\dagger$ 
and $a, b, c$ are creating and annihilation operators, respectively, and 
$\bm{\delta}$ is a vector in the physical space with $|\bm{\delta}|=a$. Expanding 
$\bm{\delta}$ in the basis $(a_\mathbf{k}, b_\mathbf{k}, c_\mathbf{k})$ leads to 
the Hamiltonian:  
\begin{align}
H_\text{Dice}=-t\begin{pmatrix}
0 & \Delta_\mathbf{k} & 0\\
\Delta^*_\mathbf{k} & 0 & \Delta_\mathbf{k} \\
0 & \Delta^*_\mathbf{k} & 0
\end{pmatrix}
\end{align}
where $\Delta_\mathbf{k}=2\exp(-ik_xa/2)\cos(\sqrt{3}/2k_ya)+\exp(ik_xa)$. The 
energy eigenvalues are $E_\pm=\pm t\sqrt{\Delta_\mathbf{k}\Delta_\mathbf{k}^*}$ 
and $E_0=0$. Figure~\ref{fig:S2_figure1}(b) shows the first Brillouin zone. At 
each of the six corners, the conduction and valence bands touch each other at 
the Dirac points $\mathbf{K}$ and $\mathbf{K}'$ (corresponding to two distinct 
valleys), which are given by
\begin{align}
\mathbf{K}=\left(\frac{2\pi}{3a}, \frac{2\pi}{3\sqrt{3}a} \right), \mathbf{K}'=\left( \frac{2\pi}{3a}, -\frac{2\pi}{3\sqrt{3}a}\right),
\end{align}
and there is a flat band through the Dirac point, as shown in 
Fig.~\ref{fig:S2_figure1}(c). For low energy excitations, the momentum relative 
to a Dirac point is $\mathbf{q}=\mathbf{k}-\mathbf{K}$, so one can write 
$\Delta_\mathbf{k} \equiv \Delta_\mathbf{q}$. Expanding $\Delta_\mathbf{q}$ about
$\mathbf{K}$ gives $\Delta_\mathbf{q} \approx q_x-iq_y$ so the energy becomes
$E_\pm\approx \pm v_F|q|$. The effective Hamiltonian for low energy excitations 
can then be written as 
\begin{align} \label{eq:Effective_H}
H = v_F \mathbf{S}\cdot \mathbf{q}, 
\end{align}
where
\begin{align} \nonumber
S_x=\frac{1}{\sqrt{2}}\begin{pmatrix}
0 & 1 & 0\\
1 & 0 & 1\\
0 & 1 & 0
\end{pmatrix}, S_y=\frac{1}{\sqrt{2}}\begin{pmatrix}
0 & -i & 0\\
i & 0 & -i\\
0 & i & 0
\end{pmatrix},
\end{align}
$\mathbf{S}$ follows general type of Levi-Civita symbols, and $v_F$ is the Fermi 
velocity. A similar energy expansion can be carried for the other valley 
$\mathbf{K}'$, leading to $H_{\mathbf{K}'}=v_F(S_xq_x-S_yq_y)$. For the dice 
lattice, there exist two distinct valleys - each triply degenerated.

\subsection{Lieb lattice}\label{subsec:Lieb}

The Lieb lattice originated the Lieb theorems of ferromagnetism for certain lattice
structures~\cite{Lieb:1989}. The first explicit lattice structure was studied
in Ref.~[\onlinecite{Tasaki:2008}]. Subsequently, a photonic 
crystal realization of the Lieb lattice was proposed\cite{SSWX:2010,AHM:2010}.
The unit cell of the Lieb lattice has three nonequivalent atoms forming a 
square-like structure, as shown in Fig.~\ref{fig:3_lattices}(c). The tight binding 
Hamiltonian is
\begin{align}
\mathcal{H}_\text{Lieb}=-2t\begin{pmatrix}
0 & \cos(k_xa/2) & \cos(k_ya/2)\\
\cos(k_xa/2) & 0 & 0 \\
\cos(k_ya/2) & 0 & 0
\end{pmatrix}.
\end{align}
The eigenvalues are $E=\pm t\sqrt{\cos^2(k_xa/2)+\cos^2(k_ya/2)}$ and $E=0$. In 
the momentum space, the conduction and valence bands touch each other at 
$k_x,k_y=\pm \pi/a$, corresponding to four Dirac points, as shown in 
Fig.~\ref{fig:3_lattices}(d). The effective Hamiltonian for the Lieb 
lattice~\cite{NOA:2013} has the same form as that for the dice lattice given by 
Eq.~(\ref{eq:Effective_H}) through a unitary transform~\cite{WXL:2021}. The 
difference is that, for a dice lattice there are two distinct valleys but there 
is only one valley for the Lieb lattice.

The Lieb lattice is experimentally accessible~\cite{leykam2018artificial} through 
fabrication techniques such as laser writing of optical 
waveguides~\cite{TOINNT:2015,VCMRMWSM:2015,MSCGOAT:2015,DLKDD:2016}. The 
electronic structure through the use of CO and Cu(1,1,1) molecules was 
studied~\cite{slot2017experimental}.  
A first-principle calculation revealed that the synthesized 2D 
$sp^2$ carbon-conjugated covalent-organic framework ($sp^2c$-COF) can have the band 
structure similar to that of the Lieb lattice~\cite{JHL:2019}.

\subsection{Kagome lattice}\label{subsec:Kagome}

The Kagome lattice originated from the study of antiferromagnet in decorated 
honeycomb lattice~\cite{Syozi:1951}. The lattice structure has the same woven 
Kagome pattern~\cite{Mekata:2003}, as shown in Fig.~\ref{fig:3_lattices}(e). 
The Kagome lattice represents a prototypical system for topological 
insulators~\cite{GF:2009,TMW:2011,CHFSBNL:2015}. The unit cell has three 
nonequivalent atoms. The tight-binding Hamiltonian is
\begin{align}
\mathcal{H}_\text{Kagome}=-2t\begin{pmatrix}
0 & \cos k_1 & \cos k_2 \\
\cos k_1 & 0 & \cos k_3 \\
\cos k_2 & \cos k_3 & 0
\end{pmatrix}
\end{align}
where $k_i=\mathbf{k}\cdot \bm{\delta}$, 
$\delta_1=\hat{x} a$, $\delta_2=(\hat{x}+\sqrt{3}\hat{y}) a/2$ and 
$\delta_3=\delta_2-\delta_1$. The energy bands are given by 
$E(\mathbf{k})=t[-1\pm\sqrt{4\Delta_\mathbf{k}-3}], 2t$ with 
$\Delta_\mathbf{k}=\cos^2k_1+\cos^2k_2+\cos^2k_3$, where the first two bands 
touch each other at six Dirac points at $E=-t$ and the third band is at $E=2t$
and is flat. The flat band thus appears at the top of the conduction band, 
as shown in Fig.~\ref{fig:3_lattices}(f). This feature is distinct from the 
dice and Lieb lattices where the flat band is located at the Dirac points.

A possible realization of the Kagome lattice through 
$\text{Ni}_3 \text{C}_{12} \text{S}_{12}$ was proposed earlier~\cite{WSL:2013}, 
where a first-principle calculation was carried out, demonstrating that nontrivial 
topological states exist in both Dirac and flat bands. In a recent 
work~\cite{Panetal:2023}, an experimental flat-band system through a self-assembled 
monolayer of 2D hydrogen-bond organic frameworks of 1,3,5-tris(4- hydroxyphenyl) 
benzene (THPB) on Au(111) surface was reported. The measured band structure fits 
well with that of the breathing-Kagome lattice. In addition, flat-to-flat band 
transitions in a diatomic Kagome lattice were reported~\cite{ZSZNL:2020}, where the
interband optical absorption coefficient exhibits a sharp peak at the gap energy, 
indicating a transition between the two flat bands. The results were further 
confirmed by first principle calculations for the material
$\text{Li}_{12}$-$(\text{Ni}_3\text{C}_{12}\text{S}_{12})_2$. Other phenomena in the
Kagome lattice include the excited quantum anomalous Hall effect~\cite{ZSLWL:2022},
an excitonic insulator~\cite{SZZYL:2021}, and theoretically proposed excitonic 
Bose-Einstein condensation~\cite{SCL:2023}.

\subsection{$\alpha$-$\mathcal{T}_3$ lattice} \label{subsec:alpha_T3}

In addition to the dice, Lieb, and Kagome lattices, another lattice structure
that can generate a flat band through the Dirac points is $\alpha$-$\mathcal{T}_3$
lattice, which is formed by adding an additional atom to the unit cell of the 
graphene lattice with tunable coupling to the nearest atoms in the original 
honeycomb lattice~\cite{Raoux2014}. The two limiting cases $\alpha = 0,1$ 
correspond to graphene and dice lattice, respectively. The tight-binding Hamiltonian
is
\begin{align}
H_{\alpha-\mathcal{T}_3}=-t\begin{pmatrix}
0 & \cos\phi\Delta_\mathbf{k} & 0\\
\cos\phi\Delta^*_\mathbf{k} & 0 & \sin\phi\Delta_\mathbf{k} \\
0 & \sin\phi\Delta^*_\mathbf{k} & 0
\end{pmatrix},
\end{align}
where $\Delta_\mathbf{k}=2\exp(-ik_xa/2)\cos(\sqrt{3}/2k_ya)+\exp(ik_xa)$
and $\alpha \equiv \tan(\phi)$ for $0 \le \phi \le \pi/4$. The effective 
Hamiltonian is
\begin{align} \label{eq:H_eff_alpha_T3}
H=\begin{pmatrix}
0 & f_k\cos(\phi) & 0 \\
f_k^* \cos(\phi) & 0 & f_k \sin(\phi)\\
0 & f_k^* \sin(\phi) & 0
\end{pmatrix},
\end{align}
where $f_\mathbf{k}=sk_x+ik_y$ and $s$ is the valley index. For $s=1$ there are 
three bands: $\tau=0,\pm 1$, corresponding to a flat band, the conduction and 
valence bands, respectively. The eigenfunctions are
\begin{align} \label{eq:aT3_eigenfunctions}
	|\psi_{\tau=\pm 1}\rangle = \frac{1}{\sqrt{2}} \begin{pmatrix} (\cos{\phi}) e^{i\theta_\mathbf{k}} \\ \tau \\
	(\sin{\phi}) e^{-\theta_\mathbf{k}}\end{pmatrix}, \ 
	|\psi_{\tau=0}\rangle = \begin{pmatrix} (\sin{\phi}) e^{i\theta_\mathbf{k}} \\ 0 \\
        -(\cos{\phi}) e^{-i\theta_\mathbf{k}} \end{pmatrix},
\end{align}
where $\theta_\mathbf{k}$ is the phase angle of $f_\mathbf{k}$: 
$f_\mathbf{k}=|f_\mathbf{k}|e^{i\theta_\mathbf{k}}$. For the other valley,
one has $f_{\mathbf{k},s=-1}=-f_{\mathbf{k},s=1}^*$, so the solution can be
obtained from a simple sign change: $\theta_\mathbf{k}=-\theta_\mathbf{k}$.

For materials such as $\text{Hg}_{1-x}\text{Cd}_{x}\text{Te}$, theoretical 
computation~\cite{MN:2015} revealed their equivalence to the 
$\alpha$-$\mathcal{T}_3$ lattice with $\alpha=1/\sqrt{3}\approx0.58$.
Experimental realizations of $\alpha$-$\mathcal{T}_3$ lattices have been 
achieved~\cite{Teppe:2016,Charnukha:2019,Hubmann:2020}.

\subsection{Additional lattices with a flat band}\label{subsec:other}

Besides 2D lattices, a flat band can also arise in 1D lattices, which 
was experimentally demonstrated using a waveguide array to simulate the atomic 
interaction~\cite{Tasaki:2008}. Observation of localized flat-band modes was made 
in a quasi-1D photonic rhombic lattice~\cite{MT:2015}. A flat band 
can also arise in 3D lattices~\cite{Orlita:2014,GM:2020}, e.g., 
in lattices with a diamond structure~\cite{NG:2005}, where the transport behavior 
in the presence of impurities was studied~\cite{GNM:2006}. The Lieb lattice can 
be extended to three dimensions, leading to the Perovskite lattice~\cite{WF:2010} 
with band gap opening. A flat band can also occur in 3D Dirac 
semimetals~\cite{YZTKMR:2012}. 
A tight-binding model for a 3D pyrochlore lattice was studied, revealing unusual flat 
band and also a flat-band enabled Weyl state~\cite{ZJHWL:2019}. The theoretical 
predictions were verified by first-principle calculations based on 
$\text{Sn}_2\text{Nb}_2\text{O}_7$.

\section{Optical properties of 2D flat-band Dirac-Weyl materials} \label{sec:optical_conductivity}

The main motivation to investigate the optical properties of Dirac-Weyl materials 
with a flat band is that the flat band offers new possibilities for electronic 
transition, so the optical conductivity could be significantly enhanced as 
compared with graphene, making these flat-band materials better candidates for 
optical sensors and modulators. For example, it was demonstrated that, when 
an external electrical field is applied to a pseudospin-1 material, the induced 
current can be two times larger than that in graphene under nonequilibrium 
conditions~\cite{WXHL:2017}, and the enhancement occurs in optical and 
magneto-optical conductivity~\cite{illes2015hall,chen2019enhanced}. Due to 
the complications brought upon by the flat band, some existing studies
focused only on the real part of the optical conductivity~\cite{illes2015hall,
tabert2016optical,kovacs2017frequency,chen2019enhanced}, leaving the crucial
issue of optical absorption largely unaddressed. A recent work~\cite{HL:2022} 
filled this gap by deriving the full optical conductivity with both real and 
imaginary parts for the $\alpha$-$\mathcal{T}_3$ lattice using the Kubo
formula~\cite{falkovsky2007space,illes2015hall}. Alternatively, the formulas
were derived~\cite{HL:2022} using the Kramers-Kronig 
method~\cite{stauber2013optical}.

There are three possible types of electronic transitions. For incident wave of
relatively low frequency $\hbar\omega<\mu$, the intraband process dominates. For 
high frequency: $\hbar\omega>\mu$, two processes become important: the transition 
from the flat band to the Dirac cone and the cone-to-cone transition, where the
former can be enhanced by increasing the value of $\alpha$, e.g., the transition 
rate for pseudospin-1 materials can be twice as large as that in 
graphene~\cite{WXHL:2017}. For the cone-to-cone transition, its rate is reduced 
with increasing $\alpha$ and becomes zero for $\alpha=1$. The complete formulas 
of the optical conductivity are general because it does not depend on other material 
properties such as the Fermi velocity~\cite{HL:2022}.

The starting point was to derive the optical matrix elements for the 
$\alpha$-$\mathcal{T}_3$ lattice. From the effective Hamiltonian 
(\ref{eq:H_eff_alpha_T3}), the current along the $x$ direction is $j_x=-ev_FS_x$, 
where
\begin{align} \nonumber
S_x=\begin{pmatrix}
0 & \cos \phi & 0\\
\cos\phi & 0 & \sin \phi \\
0 & \sin \phi & 0
\end{pmatrix}.
\end{align}
The matrix representation for the current operator is the optical matrix.
The form of the eigenfunctions in Eq.~(\ref{eq:aT3_eigenfunctions}) indicate
that, for $\mathbf{k}\ne \mathbf{k}'$, the expectation value of the current
is zero~\cite{voon1993tight}. For $\mathbf{k}=\mathbf{k}'$, one 
gets~\cite{illes2015hall}
\begin{align} \nonumber 
%\label{eq:2_matrix}
%\begin{split}
|\langle \mathbf{k}, \tau=\pm | j_x| \mathbf{k}, \tau=\pm \rangle |^2 &=e^2v_F^2\cos^2\theta_\mathbf{k}, \\ \nonumber
|\langle \mathbf{k}, \tau=\pm |j_x| \mathbf{k}, \tau=\mp \rangle |^2&=e^2v_F^2\sin^2\theta_\mathbf{k}\cos^2(2\phi), \\ \nonumber
	|\langle \mathbf{k}, \tau=0 |j_x|\mathbf{k}, \tau=\pm\rangle |^2 &=|\langle \mathbf{k}, \tau=\pm |j_x|\mathbf{k}, \tau=0\rangle |^2 \\ \label{eq:2_matrix}
	&= \frac{e^2v_F^2}{2}\sin^2\theta_\mathbf{k}\sin^2(2\phi).
%\end{split}
\end{align}
The general Kubo conductivity is 
\begin{widetext}
\begin{align} \label{eq:2_Kubo}
\sigma_{ij}(\omega,\phi)=\frac{\hbar}{2i\pi^2}\sum_{n,m}\frac{f(E_m)-f(E_n)}{E_n-E_m}\left(\frac{\langle n | j_i|m\rangle\langle m| j_j|n\rangle}{E_n-E_m-\hbar\omega}+\frac{\langle m | j_j|n\rangle\langle n | j_i|m\rangle}{E_m-E_n-\hbar\omega}\right),
\end{align}
\end{widetext}
where the subscripts $i$ and $j$ specify the directions of the current and of
the electric field, respectively. For a homogeneous material and in the absence
of any magnetic field, one has $\sigma_{xx}=\sigma_{yy}$ and
$\sigma_{xy}=\sigma_{yx}=0$. For simplicity, consider the case of $i=j=x$.
The summation is for all the state with $|n\rangle=|\mathbf{k},\tau\rangle$
and $|m\rangle=|\mathbf{k}',\tau'\rangle$. The quantity $f(E)$ in
Eq.~(\ref{eq:2_Kubo}) is the Fermi-Dirac distribution function with a
positive chemical potential $\mu$.

Due to momentum conservation, the transitions from $|n\rangle$ and $|m\rangle$
are those among the energy bands. Let $\sigma^{(1)}(\omega,\phi)$,
$\sigma^{(2)}(\omega,\phi)$ and $\sigma^{(3)}(\omega,\phi)$ denote the conductivity
due to intraband, cone-to-cone and flat-to-cone transitions, respectively.
For the intraband process, the transition is from the conduction band to
itself with $E_n-E_m\rightarrow 0$ and $E_n\approx E_m \approx \mu$, leading to
\begin{align} \nonumber
\frac{f(E_m)-f(E_n)}{E_n-E_m}=-\left. \frac{\partial f}{\partial \epsilon}\right|_{\epsilon=\mu}=\delta(\epsilon-\mu),
\end{align}
so Eq.~\eqref{eq:2_Kubo} becomes
\begin{equation} \label{eq:2_Intra1}
\sigma^{(1)}(\omega,\phi) =\frac{\hbar}{i\pi^2}\iint dk_xdk_y\frac{\partial f}{ \partial \epsilon} \frac{j_{nm}^2}{\hbar\omega}.
\end{equation}
Inserting the optical matrix element in Eq.~\eqref{eq:2_matrix} into
Eq.~(\ref{eq:2_Intra1}) and making use of the linear dispersion relationship
$E=\hbar v_F|\mathbf{k}|$, in the polar coordinates, one gets 
\begin{equation} \label{eq:2_coordinate}
\iint dk_xdk_y j_{nm}^2=\frac{e^2}{\hbar^2}\int_0^\infty \epsilon d\epsilon \int_0^{2\pi} \cos^2\theta_\mathbf{k} d\theta.
\end{equation}
Equation~\eqref{eq:2_Intra1} becomes
\begin{equation}
\sigma^{(1)}(\omega,\phi)=\frac{e^2}{i\pi\hbar^2\omega}\int \epsilon [-\delta(\epsilon-\mu)] d\epsilon=\frac{ie^2\mu}{\pi\hbar^2\omega}.
\end{equation}
With the notation $\sigma_0 \equiv e^2/(4\hbar)$, the intraband conductivity is
\begin{align} \label{eq:sigma_1}
\sigma^{(1)}(\omega,\phi)=\frac{4i\mu\sigma_0}{\pi\hbar\omega},
\end{align}
which is identical to the formula in other 2D materials such as graphene. The
denominator indicates that the intraband conductivity dominates for small
frequencies.

The cone-to-cone transitions can then be treated: those from $|\tau=-\rangle$ to 
$|\tau=+\rangle$ or vice visa (so there is an additional factor of two in the 
summation), leading to
\begin{align} \nonumber
\sigma^{(2)} (\omega,\phi)=\frac{\hbar}{i\pi^2}\sum_{n,m}\frac{f(E_m)-f(E_n)}{E_n-E_m}\frac{j_{nm}^2(-2\hbar\omega)}{(\hbar\omega)-(E_n-E_m)^2}.
\end{align}
For $\mathbf{k}=\mathbf{k}'$ and $E_n$ and $E_m$ belonging to different bands,
one can write $E_n=\epsilon$ and $E_m=-\epsilon$. Using the integral from
Eq.~\eqref{eq:2_coordinate} and the optical matrix elements Eq.~\eqref{eq:2_matrix},
one has
\begin{align} \nonumber 
%\label{eq:2_Inter1_int}
\sigma^{(2)}(\omega,\phi)=\cos^2(2\phi)\frac{e^2}{i\pi\hbar}\int [f(-\epsilon)-f(\epsilon)]\frac{\hbar\omega}{4\epsilon^2-(\hbar\omega)} d\epsilon.
\end{align}
The Fermi-Dirac distribution implies nontrivial values of
$\sigma^{(2)}(\omega,\phi)$ arise only for $\epsilon>\mu$ or $\epsilon<-\mu$
where, in the polar coordinates, only the first case contributes. This leads to
\begin{align} \nonumber
\sigma^{(2)}(\omega,\phi)=\cos^2(2\phi)\frac{e^2}{i\pi\hbar}\int_\mu^\infty \frac{\hbar\omega}{4\epsilon^2-(\hbar\omega)^2} d\epsilon .
\end{align}
The integral has a singularity for $2\hbar\omega>\mu$. Using the residue
theorem, one gets
\begin{align} \label{eq:sigma_2}
\sigma^{(2)}(\omega,\phi)=\cos^2(2\phi)\sigma_0\left[\Theta(\hbar\omega-2\mu)-\frac{i}{\pi}\ln \left|\frac{\hbar\omega+2\mu}{\hbar\omega-2\mu} \right| \right],
\end{align}
where $\Theta$ is the Heaviside step function. It can be verified that, for $\phi=0$, 
the result coincides with that for graphene. At the opposite end of the 
$\alpha$-$\mathcal{T}_3$ spectrum $\phi=\pi/4$ (pseudospin-1), the integral is zero.

The same method can be used to obtain the contribution of the flat-to-cone
transitions to the optical conductivity. In this case, $E_n=0$ and
$E_m=\epsilon$, so
\begin{align} \nonumber
\sigma^{(3)}(\omega,\phi)=\sin^2(2\phi)\frac{e^2}{i\pi\hbar}\int_\mu^\infty \frac{\hbar\omega}{\epsilon^2-(\hbar\omega)^2} d\epsilon,
\end{align}
where the singularity occurs at $\hbar\omega=\epsilon$ and the weight becomes
$\sin^2(2\phi)$. Evaluating this integral gives
\begin{align} \label{eq:sigma_3}
\sigma^{(3)}(\omega,\phi)=2\sin^2(2\phi)\sigma_0\left[\Theta(\hbar\omega-\mu)-\frac{i}{\pi}\ln \left|\frac{\hbar\omega+\mu}{\hbar\omega-\mu} \right| \right].
\end{align}

\begin{figure} [ht!]
\centering
\includegraphics[width=\linewidth]{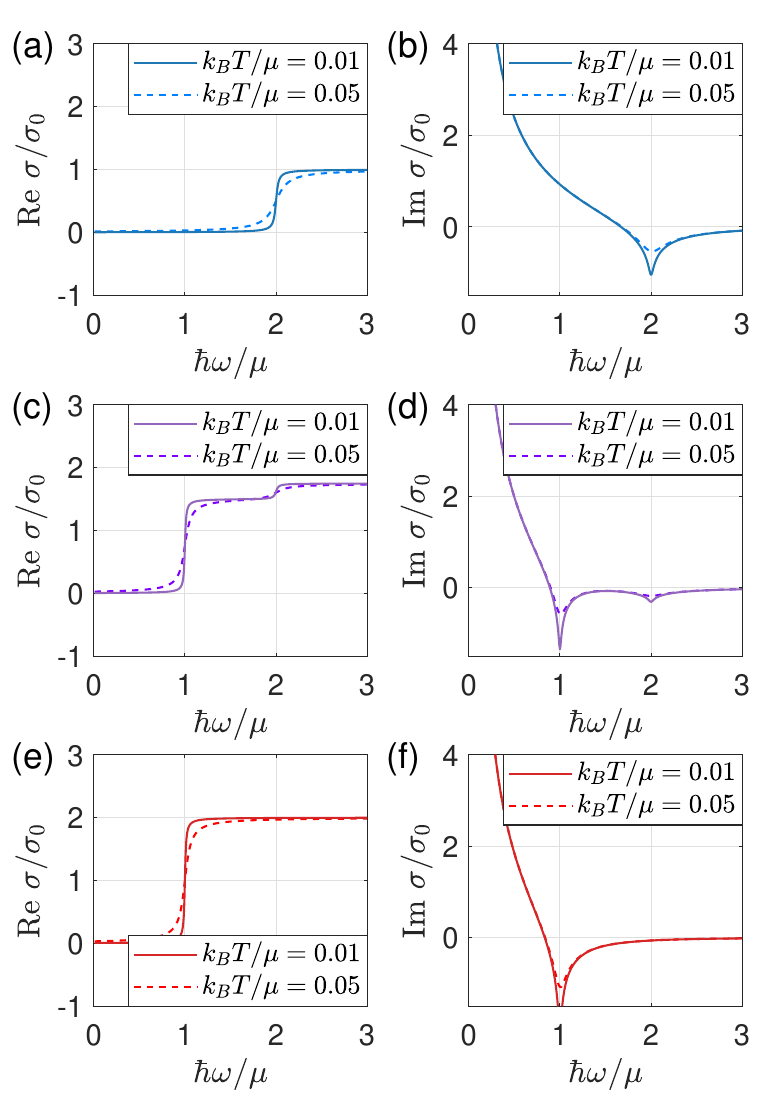}
\caption{Real and imaginary parts of the optical conductivity of the 
$\alpha$-$\mathcal{T}_3$ lattice derived from the Kubo formula in the absence of 
any impurity scattering. (a,c,e) Real part of the optical conductivity for 
$\alpha=0$ (graphene), $\alpha=1/\sqrt{3}$, and $\alpha=1$ (pseudospin-1), 
respectively. At zero temperature, the conductivity is nonzero for
$\hbar\omega/\mu>2$. An interband transition leads to a dip in the conductivity 
plot. The step-function type of transition is smoothed out by finite temperatures. 
(b,d,f) Imaginary part of the optical conductivity for $\alpha=0$, $1/\sqrt{3}$, 
and 1, respectively. Reprinted with permission from [C.-D. Han and Y.-C. Lai, 
``Optical response of two-dimensional Dirac materials with a flat band,'' Phys. 
Rev. B 105, 155405 (2022)]. Copyright (2022) by the American Physical Society.}
\label{fig:Kubo}
\end{figure}

These conductivity formulas suggest that Dirac-Weyl flat-band materials
can have significantly larger optical conductivity than that for graphene,
due to the much stronger interaction between light and the lattice structure
of the materials~\cite{HL:2022}. As an example, Figs.~\ref{fig:Kubo}(a),
\ref{fig:Kubo}(c) and \ref{fig:Kubo}(e) show the real part of the 
finite-temperature optical conductivity for three different values of
$\alpha$, respectively, to which the intraband process has no contribution.
For $\alpha = 0$ [graphene, Fig.~\ref{fig:Kubo}(a)], only the cone-to-cone
transition exists. For $\alpha =1/\sqrt{3}$ [Fig.~\ref{fig:Kubo}(c)], there
are two transition points: cone-to-cone transition for $\hbar\omega/\mu>2$ 
and flat-band-to-cone transition for $\hbar\omega/\mu>1$. For $\alpha =1$ 
[Fig.~\ref{fig:Kubo}(e)], flat-band-to-cone transition is the only possibility
and its magnitude is twice of that of the cone-to-cone transition for 
graphene. The respective imaginary parts of the conductivity are shown in 
Figs.~\ref{fig:Kubo}(b), \ref{fig:Kubo}(d), and \ref{fig:Kubo}(f). In all
three cases, the intraband process gives a singularity at $\omega\rightarrow 0$, 
and each interband transition leads to a dip for Im $(\sigma) < 0$. Note
that the imaginary part of the conductivity can be negative. Previously, 
it was found for graphene that a negative imaginary part can lead to
a special TE mode for electromagnetic wave propagation~\cite{mikhailov2007new}. 
For the $\alpha$-$\mathcal{T}_3$ lattice, a negative imaginary part of
the conductivity can have a significant effect on the intrinsic plasmon modes
with respect to the loss, confinement and impurity scattering~\cite{HL:2022}.

The formulas Eqs.~(\ref{eq:sigma_1}), (\ref{eq:sigma_2}), and (\ref{eq:sigma_3})
give a complete description of the optical conductivity of the 
$\alpha$-$\mathcal{T}_3$ lattice, which were verified~\cite{HL:2022} by an 
independent theoretical approach: the Kramers-Kronig formula. As examples of the
application of the conductivity formulas, two phenomena were studied~\cite{HL:2022}. 
First, while intraband transition leads to TM polarized waves at low frequencies 
(1–10 THz), TE polarized waves can emerge at high frequencies (100–300 THz), due 
to the two interband transitions. Second, the unique flat-band-to-cone transition 
generates multifrequency TE propagating waves and a strong optical response. These 
phenomena were numerically confirmed by the behaviors of propagating surface wave 
and scattering~\cite{HL:2022}.

\section{Discussion} \label{sec:discussion}

In general, the optical responses of flat-band Dirac-Weyl materials are stronger
than those of graphene, as conductivity due to the flat-band-to-cone transition is 
twice of that induced by cone-to-cone transition. The physical reason behind is 
that the plane waves in these materials have a smaller attenuation length due to the 
large imaginary part of the optical conductivity as compared to that in graphene. 
This means that, at the same frequency, a larger scattering cross section can arise
in flat-band Dirac-Weyl materials.

A complete description of the optical conductivity of flat-band Dirac-Weyl materials 
opens the door to investigating problems pertinent to development of optical devices. 
For example, intrinsic plasmon modes whose physical properties depend on the 
polarization were studied~\cite{HL:2022} with the finding that TM waves are the 
result of intraband transitions, which usually occur in the frequency range 1--10 THz,
but TE waves are the result of interband transitions, which can arise in a higher 
frequency range: 100--300 THz. When two interband transitions occurs (e.g., for 
$0 < \alpha < 1$ in the $\alpha$-$\mathcal{T}_3$ lattice), two TE surface waves can 
arise, respectively, at $\hbar\omega/\mu\approx 1, 2$. It was also suggested that TE 
polarized waves can be tuned by adjusting the chemical potential~\cite{HL:2022}. 
Another example is scattering from a dielectric sphere coated with multiple layers 
of flat-band Dirac-Weyl material~\cite{HL:2022}, where TM wave scattering can be 
stronger than TE wave, due to the reduced imaginary part of the optical conductivity 
at finite temperatures. This phenomenon can be exploited for enhancing certain 
desired polarization. A full optical conductivity theory allows electromagnetic 
dynamics in flat-band Dirac-Weyl materials to be studied in detail. Issues such as 
the emergence of intrinsic plasmon modes at a single or multiple frequencies, loss, 
attenuation length, and finite temperatures can be studied in detail. In fact, the 
occurrence of multi-frequency plasmon modes implies the possibility of achieving
superscattering or cloaking at multiple frequencies. These broadband effects
can find applications in optical sensing, imaging, tagging or 
spectroscopy~\cite{danilov2000detection,del2011octave,schliesser2012mid}.
Moreover, edge states in graphene can lead to a blue shift in the plasmon 
modes~\cite{wedel2018emergent}. To exploit flat-band Dirac-Weyl materials for 
applications in quantum plasmonics could be an interesting area of research.

We briefly discuss the effects of impurities and many-body interaction on the optical 
response.

\paragraph*{Effects of impurities.}
A number of previous works addressed this issue, but mainly for graphene. The
general methodology is to start from the linear dispersion relationship and model
the effects of defects or impurities on optical scattering through the incorporation
of a relaxation time, taking into account electron-phonon scattering. For graphene,
the relaxation time is relatively small, so it affects the low-frequency response
more than the high-frequency response, rendering negligible the effect on optical
response~\cite{JBS:2009}. In another work that
went beyond the Dirac-cone approximation~\cite{YRDK:2011}, the
authors used the tight-binding model and the Kubo formula to study the effects of
different types of impurities in graphene on the optical response, which included
random potentials, random vacancies and random coupling, and found that the
interband transition strength decreases with the impurity density. For example,
for lattice vacancies, the interband transition is strong for 5\% of the vacancies
but is barely observable for 10\% of the vacancies. For general types of impurities,
their effects on the transition cannot be neglected. Since the impurities can
generate states at $E=0$, in graphene with defects, a transition at
$\hbar\omega=\mu$ can occur. In a more recent work~\cite{VWKF:2017}, 
hydrogen atoms as impurities were added to a graphene sheet at the
density of approximately 300 impurity atoms per $\mu m^2$. For $\mu=2 {\rm eV}$, 
these impurities have negligible effect on the optical conductivity. However, for
$\mu=0.2 {\rm eV}$, an observable dip in the conductivity occurs at 
$\hbar\omega=2\mu$.

For 2D pseudospin-1 Dirac materials with a flat band, the effects of impurities on
optical response can be treated similarly by incorporating a finite relaxation time
into the $\alpha$-$T_3$ lattice~\cite{HL:2022}. Under the same impurity conditions, 
for $\alpha=1$ the interband transition is two times stronger than that in graphene, 
so this transition is more robust against defects or vacancies. It was
found that, even when the relaxation time is several times smaller than that in
graphene (corresponding to a more significant amount of impurities), the effects
on the optical response in the high frequency regime are insignificant. In a recent
work on the $\alpha$-$T_3$ lattice~\cite{LWL:2023}, the
effects of lattice vacancies (up to 0.4\%) leading to different inelastic-scattering
rates on the density of states were studied and found to be negligible. However, the
inelastic scattering can lead to a broadening of the flat band.
In another recent work~\cite{IZGH:2023}, the optical
conductivity in $\alpha$-$T_3$ lattice with a distorted flat band was studied
and results similar to those in Ref.~[\onlinecite{HL:2022}] were
found, including the dependence of the conductivity on the temperature.

\paragraph*{Effects of many-body interactions on optical response.}
In graphene, the electron self energy was used to describe the electron-electron
interaction~\cite{GVV:2009}, and the scattering conductivity results were compared 
with the experimental measurements, validating the approach~\cite{LHJHMKSB:2008}.
The electron-electron interactions and impurity scattering can reduce the transition 
strength by $20\%$. An increase in the real part of the conductivity at 
$\hbar\omega=2\mu$ was observed for a wide interval of $\mu$. It was also observed 
that the many-body effect and impurities in graphene create a non-zero optical 
conductivity for energy less than $\hbar\omega=2\mu$, with the transition strength 
about $80\%$ of that of the clean lattice. For 2D Dirac materials with a flat band, 
the combined lattice impurities and many-body interactions in general will lead to 
a reduction in the optical transition strength as compared with graphene, but the
issue remains to be outstanding.

\section*{Acknowledgement}

This work was supported by AFOSR under Grant No.~FA9550-21-1-0186.

\bibliography{Flatband}

\end{document}